\documentclass[proof]{WileyASNA-v1}
\usepackage{hyperref}

\articletype{ORIGINAL ARTICLE}%

\received{}
\revised{}
\accepted{}

\begin{document}

\title{Limitations of the modified blackbody fit method for determining molecular cloud properties}

\author[1]{N. Zielinski*}

\author[1]{S. Wolf}

\authormark{Zielinski \& Wolf}

\address[]{ University
 of Kiel, Institute of Theoretical Physics and Astrophysics, Leibnizstrasse 15, 24118 Kiel, Germany}

\corres{*Niko Zielinski, \email{nzielinski@astrophysik.uni-kiel.de}}


\abstract{ Achieving a comprehensive understanding of the star and planet formation process is one of the fundamental tasks of astrophysics, requiring detailed knowledge of the physical conditions during the different phases of this process.
During the earliest stages, i.e., concerning physical processes in molecular clouds and filaments, the column density $N$(H$_2)$, dust temperature $T$ and dust emissivity index $\beta$ of these objects can be derived by adopting a modified blackbody fit of the far-infrared to (sub-)millimeter spectral energy distributions.
However, this often applied method is based on various assumptions. In addition, the observational basis and required, but only assumed cloud properties, such as a limited wavelength-coverage of the spectral energy distribution and dust properties, respectively, may differ between different studies.
 We review the basic limitations of this method and evaluate their impact on the derived physical properties of the objects of interest, i.e., molecular clouds and filaments. 
We find that the highest uncertainty when applying this method is introduced by the often poorly constrained dust properties. Therefore, we propose to first derive the optical depth and subsequently the column density with the help of a suitable dust model as the optical depth can be obtained with high accuracy, especially at longer wavelengths. 
The method provides reliable results up to the high densities and corresponding optical depths observed in molecular clouds. 
Considering typically used observational data, i.e., measurements obtained with far-infrared instruments like Herschel/PACS, JCMT/SCUBA-2 and SOFIA/HAWC+, data at four wavelengths are sufficient to obtain accurate results. 
Furthermore, we find that the dust emissivity index $\beta$ derived with this method is not suitable as an indicator of dust grain size. }


\jnlcitation{\cname{%
\author{Zielinski, N.},  
\author{Wolf, S.}} (\cyear{2023}), 
\ctitle{Limitations of the modified blackbody fit method for determining molecular cloud properties}, \cjournal{Astron. Nachr.}}


\maketitle

\section{Introduction} \label{Section:Introduction}

Understanding the process of star and planet formation, starting from the collapse of a dense molecular cloud core, through the formation of circumstellar disks and the onset of dust coagulation toward the formation of planets, is a longstanding challenge of modern astrophysics \citep[e.g.,][]{Lissauer1993, Testi2014, Krumholz2014, Liu2020b, Raymond2022}.
In this context it is indispensable to study the initial conditions of this process and to constrain the influence of the key physical components and effects, such as magnetic fields or turbulence \citep[e.g.,][]{Shu1987, Federrath2016, Orkisz2017, Soler2019}.
In particular, constraints on physical properties of  molecular clouds and filaments, such as the spatial distribution of the dust temperature $T$ and column density $N$(H$_2)$, are essential. These basic quantities can be derived from the spatially resolved or unresolved spectral energy distribution (SED) of these objects in the far-infrared (FIR) to millimeter (mm) wavelength range. An often applied approach in this analysis is to fit the observed SED by a modified blackbody\footnote{In short "modified blackbody fit method"  or abbreviated as BFM throughout this study.} \citep[e.g.,][]{Hildebrand1983, Dupac2002, Hill2009}. 
Moreover, this method allows one to derive the dust emissivity index $\beta$. 
Making use of observations obtained, e.g., with Herschel PACS\footnote{Photodetector Array Camera and Spectrometer}/SPIRE\footnote{Spectral and Photometric Imaging Receiver} \citep{Pilbratt2010, Poglitsch2010, Griffin2010} and SOFIA\footnote{Stratospheric Observatory for Infrared Astronomy}/HAWC+\footnote{High-resolution Airborne Wideband Camera Plus} \citep{Temi2018, Harper2019}, this method has been applied in various studies \citep[e.g.,][]{Gandilo2016, Fissel2016, Pokhrel2016, Lin2017, RiveraIngraham2017, Potdar2022, Matsuura2022, LopezRodriguez2022}. 
In particular, this method played an increasing role in studies of magnetic fields in star-forming clouds and filaments in recent years \citep[e.g.,][]{Santos2019, Chuss2019}. The structure and strength of the magnetic field in these objects can be constrained by observing the continuum polarization resulting from the emission and absorption of magnetically aligned non-spherical dust grains \citep[e.g,][]{Wolf2003, Crutcher2004, Kirk2006, Kwon2018}.
However, dust polarization can only be adopted as a reliable tracer of magnetic field structures if the underlying dust grain alignment mechanism is understood. 
There are various theories of dust grain alignment, e.g., Davis-Greenstein \citep{DavisGreenstein1951}, mechanical alignment \citep{Gold1952} or radiative alignment \citep{Dolginov1976, Lazarian2007}, with radiative alignment torque (RAT) being the most promising one. 
In order to constrain the dust grain alignment efficiency and thus to derive magnetic field properties, the density and temperature structure as well as dust grain properties have to be determined. \\
However, the modified blackbody fit method as an approach to derive these quantities has some potential weaknesses. First and foremost, any uncertainties concerning the dust properties have a direct impact on the derived cloud properties. In an attempt to mitigate this issue,  a power-law approach is applied to describe the wavelength-dependent dust opacity $\kappa_\nu$:
\begin{equation} \label{Equation:Dust_Opacity}
\kappa_\nu = \kappa_{\nu_0} \, \left( \frac{\nu}{\nu_0} \right)^\beta \: .
\end{equation}
Here, $\kappa_{\nu_0}$ is the reference opacity at a reference frequency $\nu_0$.
 The origin of this solution can be traced back to \citet{Hildebrand1983} and the dust opacity stated there, resulting from observations of the reflection nebula NGC\,7023. \\ 
 Another potential problem inherent to this method is given by the limitation of the available observations and thus the resulting limited wavelength coverage in the FIR to sub-millimeter wavelength range. This problem becomes particularly severe when there is no observatory/instrument available covering the \mbox{$\sim$ 30-250\,$\mu$m} wavelength range, i.e., since the end of observations with the Herschel Space Observatory and SOFIA.
  However, even on the basis of archival data, e.g., if only SOFIA/HAWC+ or Herschel/PACS data are available for the object to be studied, an insufficient coverage of the Rayleigh-Jeans or Wien parts of the SED could lead to uncertainties of the derived physical parameters. A third potential problem when adopting this method arises, if the angular resolution is too low to resolve embedded regions with significantly higher density, e.g., clumps and filamentary structures.
A high optical thickness as well as a spatially variable size distribution of the dust resulting from increased dust grain growth in these denser regions \citep[e.g.,][]{Steinacker2014, Steinacker2015} could lead to further uncertainties.\\
Given the outlined potential limitations of this method, the goal of this study is to determine the impact of these limitations on the parameter estimates, i.e., that of the derived dust temperature, column density and dust emissivity index.
This paper is organized as follows: In Sect. \ref{Section:Description_of_the_fitting_process}, we introduce the modified blackbody method and present  the model we used to analyze the limitations of this method in Sect. \ref{Section:Model}. We divide the presentation of the results in two parts. First, we discuss the influence of the dust properties (Sect. \ref{Section_Influence_Dust_Properties}) and the specific choice of observing wavelengths (Sect. \ref{Section:Fitting_procedure_for_selected_wavelengths}). Second, we focus on the role of the characteristics of the source, i.e., the molecular cloud, (Sect. \ref{Section:Characteristics_of_the_source}), investigating the influence of optical depth (Sect. \ref{Section:Influence_of_the_optical_depth}), the specific dust model (Sect. \ref{Section:Different_Dust_Models}) and temperature profiles (Sect. \ref{Section:Case_of_non_constant_temperature}) on the derived quantities.
Finally, we summarize our results in Sect. \ref{Section_Conclusion}.

\section{Details about the modified blackbody fit method and the model space} \label{Section:Details_about_the_blackbody_fit_method_and_the_model_space}
In this chapter, we summarize the modified blackbody fit method (BFM, Sect. \ref{Section:Description_of_the_fitting_process}) and introduce the model we adopt for this study (Sect. \ref{Section:Model}).
\subsection{Fitting method} \label{Section:Description_of_the_fitting_process}
Using continuum observations at FIR to mm wavelengths, the BFM allows constraining the column density $N$(H$_2)$, temperature $T$ and dust emissivity index $\beta$. 
The origin of this method can be traced back to \citet{Hildebrand1983}, \citet{Dupac2001} and \citet{Vaillancourt2002}. We follow the description of the method as outlined by \citet{Chuss2019}. The intensity $I_\nu$ of a dust cloud with equilibrium temperature $T$ and optical depth $\tau(\nu)$ along the line of sight can be expressed as
\begin{equation}
I_\nu = (1 - \exp\left( -\tau(\nu) \right) \: B_\nu(T) \: .
\end{equation}
Here $B_\nu(T)$ is the Planck function for temperature $T$ at frequency $\nu$. The optical depth $\tau(\nu)$ is expressed as:
\begin{equation} \label{Formular:Optical_depth}
\tau(\nu) = \epsilon \, \left( \frac{\nu}{\nu_0} \right)^\beta \: ,
\end{equation}
where $\epsilon$ is a constant of proportionality, directly related to the column density along the line of sight, and $\beta$ is the dust emissivity index. The constant of proportionality is given by 
\begin{equation}
\epsilon = \kappa_{\nu_0} \, \mu \, m_\text{H} \, N\mathrm{(H_2)} \: .
\end{equation}
Here, $\kappa_{\nu_0}$ is a reference dust opacity per unit mass at frequency $\nu_0$, $\mu$ is the mean molecular weight per hydrogen atom, $m_\text{H}$ is the atomic mass of hydrogen, and $N$(H$_2$) is the gas column density. 
To be consistent with values applied in previous studies \citep[e.g., ][]{Koenyves2010,Santos2019, Chuss2019, Sandell2021, Fiorellino2021, LopezRodriguez2022, Matsuura2022, Azatyan2022}, we adopt \mbox{$\nu_0$ = 1000\,GHz}, $\kappa_{\nu_0}$ = 0.1 cm$^2$ g$^{-1}$ and \mbox{$\mu$ = 2.8}. For more details regarding the dust model, see Sect. \ref{Section_Influence_Dust_Properties}. 
The resulting fit function is
\begin{multline} \label{Equation:Complete_Fit_Equation}
 I_\nu = \left( 1 - \exp \left(- \kappa_{\nu_0} \: \mu \: m_H \: N(\mathrm{H}_2) \: \left(\nu/\nu_0\right)^\beta \right)  \right) \\\ \hspace{-10cm}  \frac{2h \nu^3}{c^2} \:  \frac{1}{\exp\left(\frac{h \nu}{k T}\right) -1} \, .
\end{multline}
The fitting parameters are the column density $N$(H$_2)$, the temperature $T$ and the dust emissivity index $\beta$. \\
In order to investigate the impact of the limitations described in Sect. \ref{Section:Introduction} on these derived fitting parameters, we utilize the \textit{curve$\_$fit} function (with the Levenberg-Marquardt algorithm for optimization) from the Scipy Python package \citep{Virtanen2020} for the fitting process. 
Essentially, the complete fitting process consists of a series of pixel-wise SED fits resulting from spatially resolved flux continuum observations at multiple wavelengths.
Working with real data, it is imperative to re-project the data to a common pixel scale and to beam-convolve to a common resolution prior to the fitting process. 
In the current study this preparatory step is not necessary, since we analyze synthetic data with a common pixel scale.  
Our analysis is based on simulated ideal observations at nine wavelengths: 53, 89, 154, 214$\,\mu$m (SOFIA/HAWC+ bands A, C, D, E), 70, 100, 160$\,\mu$m (Herschel/PACS), 250\,$\mu$m (Herschel/SPIRE) and 850\,$\mu$m (JCMT/SCUBA-2), see Table \ref{Table:Overview_instruments_wavelengths} for an overview of the selected observing wavelengths with the associated beam sizes and uncertainties. We do not take observations at Herschel/SPIRE 350\,$\mu$m and 500\,$\mu$m into account, since the resolution (25$''$, 36$''$) is low in comparison to the bands listed above. Measurements with JCMT/SCUBA-2 at 450\,$\mu$m are not considered either since the related uncertainties are high \citep[$\sim$ 50\%,][]{Sadavoy2013} in comparison to the selection we adopt (10-20\,\%, see Tab. \ref{Table:Overview_instruments_wavelengths}) \\
Using the 3D Monte-Carlo radiative transfer code \texttt{POLARIS}\footnote{\url{https://portia.astrophysik.uni-kiel.de/polaris/}} \citep{Reissl2016}, we calculate intensity maps based on underlying dust temperature and density distributions. Each map is then beam-convolved to a resolution of 18.2$''$, which is the lowest resolution within the selected instrument/observing wavelengths. We then perform the fitting procedure and compare the results for column density and temperature with the (original) values for these parameters. 
Applying the BFM one has to be aware, that the derived temperature distribution in the plane of sky represents only average values for the potentially variable temperature structure along the line of sight.\\
As part of this study, different dust models and their effect on the parameter estimates are considered. In addition, a distinction is made whether $\beta$ is treated as a fitting parameter or is set to a fixed value, resulting in three different cases: 
\begin{itemize}
\item[i)] "Ideal case": Application of the dust properties of our considered dust mixture for the fitting.
\item[ii)] $\beta$ fixed: Application of a power-law: $\kappa_\nu = \kappa_{\nu_0} \, \left( \frac{\nu}{\nu_0} \right)^\beta$, with \mbox{$\kappa_{\nu_0}$ = 0.1 cm$^2$ g$^{-1}$}, $\nu_0$ = 1000\,GHz and $\beta$ fixed to 1.62 \citep[e.g.,][]{PlanckCollaboration2014, Santos2019} or 2 \citep[e.g.,][]{Gandilo2016, Fissel2016}.
\item[iii)] "$\beta$ free": Power-law approach for the opacity as in ii),  but with $\beta$ as fitting parameter. 

\end{itemize}
We filter the fitting results by adopting the reduced $\chi^2_r$ as a measure of the goodness of the fit and dismiss all results with \mbox{$\chi^2_r$ $>$ 10}.

\subsection{Model} \label{Section:Model}
In this section we describe the reference model we use in this study and which will be the basis of our analysis. While this model shall allow reproducing the fitting process applied in earlier studies, it will provide the basis for an analysis of the influence of selected physical parameters and selected frequently used basic assumptions of the BFM.
 The model space is a sphere with a constant dust density at a distance of 388\,pc \citep[distance to the Orion molecular cloud;][]{Kounkel2017}. The sphere has a diameter of 6.2$'$ which is the size of the field-of-view (FOV) for SOFIA/HAWC+ band E \citep[intensity FOV of Band E is \mbox{8.4$'$ $\times$ 6.2$'$;}][]{Harper2019}, resulting in a radius of $R$ = 70604\,au. \\
For our reference model we consider a dust model which consists of a mixture of astrosilicate (abundance of 62.5\%) and graphite\footnote{According to the crystal structure of graphite its anisotropic optical properties are considered by applying the 1/3 - 2/3 approximation \citep{Draine1984, Draine1993}.} grains (abundance of 37.5\%), hereafter Sil-Graph, which was applied in numerous earlier studies \citep[e.g.,][]{Wolf2003b, Sauter2009, Das2010, Reissl2014, Siebenmorgen2014, Guillet2018, Valdivia2019, Kobus2020, LeGouellec2020, Brunngraeber2020, Lee2020, Zielinski2021a, Li2022, Fanciullo2022, Chastenet2022}. The dust grains are compact, homogeneous and of spherical shape, with optical properties from \citet{Draine2003}. The dust grain size distribution is the same as that in the ISM:
 \begin{equation}
 \mathrm{d}n(a) \propto a^{-3.5} \: \mathrm{d}a, \qquad a_\mathrm{min} < a < a_\mathrm{max}, 
 \end{equation}
where d$n(a)$ denotes the number of dust grains with a radius which is in the range $[a, a+\mathrm{d}a]$. We set the minimum/maximum dust grain size to $a_\text{min}$ = 5\,nm/$a_\text{max}$ = 250\,nm \citep{MathisRumplNordsieck1977} and adopt the canonical dust-to-gas mass ratio of 0.01.
We assume a constant dust temperature of 20\,K and a total gas mass of $M_\text{gas}$ = 5 $\cdot$ 10$^2$\,M$_\odot$. The corresponding optical depth is less than unity for each observing wavelength we consider for the fitting process (see Tab. \ref{Table:Overview_instruments_wavelengths}). The corresponding column density of \mbox{$\overline{ N(\mathrm{H}_2)}$ $\sim$ 7 $\cdot$ 10$^{22}$ cm$^{-2}$ }is within the typical range derived for molecular clouds of 10$^{20}$ -- $\sim$10$^{24}$ cm$^{-2}$ \citep[e.g.,][]{Koenyves2015, Pokhrel2016, Fissel2016, Lin2017, Santos2019, Chuss2019}.

\begin{table*} [t]
  \begin{center}
    \caption{\: Overview of instruments \& related observing wavelengths we utilize for the BFM. The uncertainties of the flux measurements are adopted from \citet{Chuss2019}.}
    \label{Table:Overview_instruments_wavelengths}
  \noindent   \begin{tabular}{cccc}
    \hline \hline    \rule{0pt}{3ex}
      Instrument & Wavelength [$\mu$m] & Beam size [$''$] & Uncertainty [\%]  \\
      \hline \hline
      \rule{0pt}{3ex}
     \noindent SOFIA/HAWC+$^a$ & 53 & 5.1 & 15 \\ \rule{0pt}{2ex}
     \noindent Herschel/PACS$^b$ & 70 & 5.6 & 20 \\ \rule{0pt}{2ex}  
     \noindent SOFIA/HAWC+$^a$ & 89 & 7.9 & 15 \\ \rule{0pt}{2ex}     
     \noindent Herschel/PACS$^b$ & 100 & 6.8 & 20 \\ \rule{0pt}{2ex}
     \noindent SOFIA/HAWC+$^a$ & 154 & 14.0 & 15 \\ \rule{0pt}{2ex}     
     \noindent Herschel/PACS$^b$ & 160 & 11.3 & 20 \\ \rule{0pt}{2ex}     
     \noindent SOFIA/HAWC+$^a$ & 214 & 18.7 & 20 \\ \rule{0pt}{2ex}     
     \noindent Herschel/SPIRE$^c$ & 250 & 18.1 & 10 \\ \rule{0pt}{2ex}
     \noindent JCMT/SCUBA-2$^d$ & 850 & 14.2 & 15 
\\      \hline \hline
    \end{tabular} 
     \\{\footnotesize $^a$\citealt{Harper2019}, $^b$\citealt{Poglitsch2010}, $^c$\citealt{Griffin2010},  $^d$\citealt{Dempsey2013}.}
  \end{center}
\end{table*}

\begin{table}
  \begin{center}
    \caption{\: Reference model. Free parameters are marked with the symbol "$\updownarrow$".}
    \label{Table:Overview_parameters_model}
  \noindent   \begin{tabular}{ccc}
    \hline \hline    \rule{0pt}{3ex}
      Parameter & Symbol & Value  \\
      \hline \hline
      \rule{0pt}{3ex}
     \noindent  Gas density  & $\rho$  & constant   \\ \rule{0pt}{2ex}
     \noindent  Radius  &  $R$  & 70604\,au   \\ \rule{0pt}{2ex}
      \noindent Gas mass  &  $M_\text{gas}$ $\updownarrow$ & 500\,M$_\odot$   \\ \rule{0pt}{2ex}
      \noindent Dust-to-gas ratio  &  $f_{d/g}$  & 0.01   \\ \rule{0pt}{2ex}
      \noindent Dust model  &    &  Sil-Graph $\updownarrow$  \\ \rule{0pt}{2ex}
      \noindent Minimum dust grain radius  &  $a_\text{min}$  &  5\,nm   \\ \rule{0pt}{2ex}
      \noindent Maximum dust grain radius  &  $a_\text{max}$ $\updownarrow$  &  250\,nm   \\ \rule{0pt}{2ex}
      \noindent Exponent of grain size distribution  &  $\alpha$  &  -3.5   \\ \rule{0pt}{2ex}
      \noindent Number of wavelengths used for fit &  $n_\lambda$ $\updownarrow$ &  9   \\ \rule{0pt}{2ex}
     \noindent  Distance  & $d$  & 388\,pc   
\\      \hline \hline
    \end{tabular} 
  \end{center}
\end{table}

\section{Results I: Technical aspects and limitations} \label{Section:Technical_aspects_and_limitations}
In the following, we discuss intrinsic problems of the BFM with regard to uncertainties of the dust properties and the selection of observing wavelengths.
The impact of specific characteristics of the source on the physical quantities derived with the BFM, such as different dust models, grain sizes or optically thick regimes, will be discussed in Sect. \ref{Section:Characteristics_of_the_source}.

\subsection{Influence of dust properties} \label{Section_Influence_Dust_Properties}
\begin{figure*} 
\includegraphics[width=\hsize]{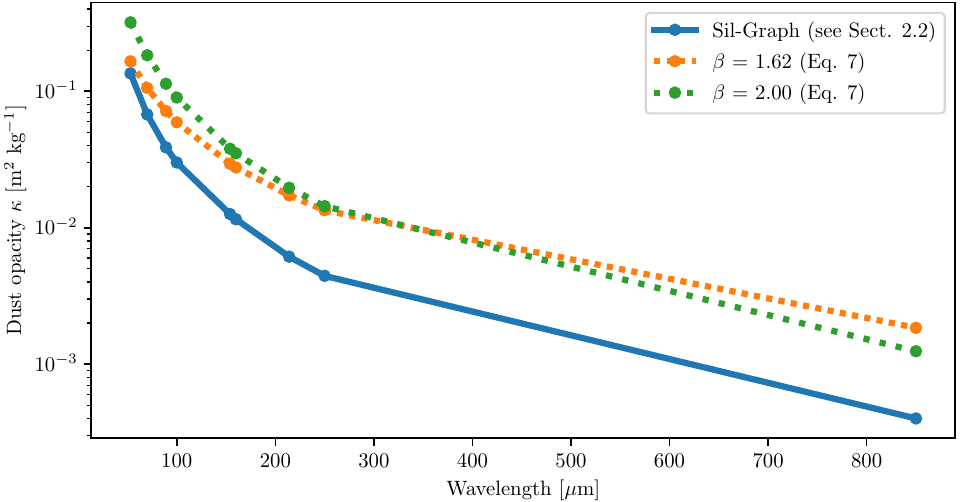} \caption{\: Wavelength-dependent dust opacity $\kappa_\nu$ for the considered dust models: Sil-Graph (see Sect. \ref{Section:Model}), power-law with \mbox{$\beta$ = 1.62}, power-law with $\beta$ = 2.0 (see Eq. \ref{Equation:Dust_Opacity}). } \label{Figure:Overview_Dust_Properties}
\end{figure*}

\begin{figure*} 
\includegraphics[width=1.0\textwidth]{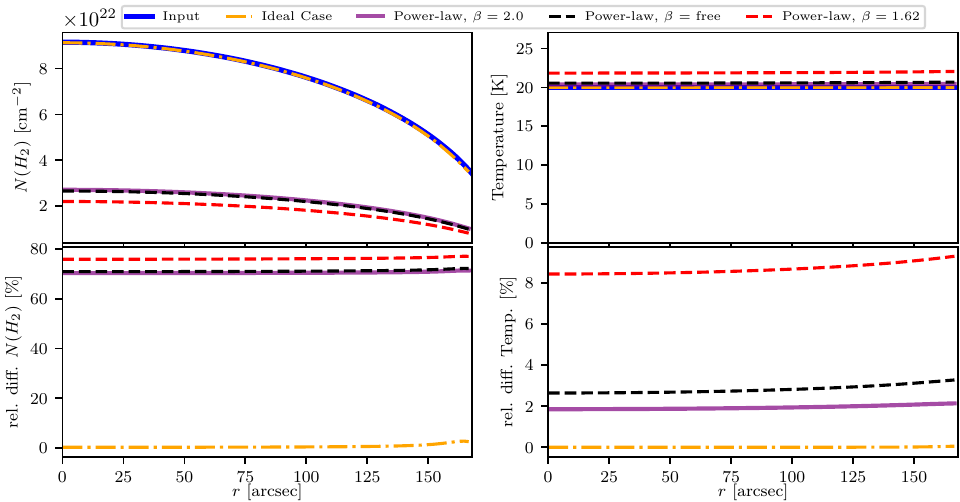} \caption{\: Results of the fitting process for four different cases: ideal case (i.e., known dust properties), power-law approach with $\beta$ = (1.62, 2) and $\beta$ as a fitting parameter (see Eq. \ref{Equation:Dust_Opacity}). \textit{Top:} Radial profiles of column density and temperature. The blue line marks the input values for the column density and temperature, respectively. \textit{Bottom:} Relative deviations between the results of the fitting process and the input values for column density (left) and temperature (right).} \label{Figure:Fit_NH2_Beta_T_5e2Msun_T_const20K_all_wl}
\end{figure*}

The first general problem inherent to the BFM is the uncertainty about the optical properties of the dust. As mentioned in Sect. \ref{Section:Introduction}, a usual approach to circumvent -- but not to solve -- this problem is to adopt a power-law, as already described in \ref{Section:Model} and Eq. \ref{Equation:Complete_Fit_Equation}, for the frequency dependence of the dust opacity:
\begin{equation} \label{Equation:Dust_Opacity}
\kappa_\nu = \kappa_{\nu_0} \, \left( \frac{\nu}{\nu_0} \right)^\beta \: .
\end{equation}
In ealier studies, values of  \mbox{$\kappa_{\nu_0}$ = 0.1 cm$^2$ g$^{-1}$} and \mbox{$\nu_0$ = 1000\,GHz} (corresponding to a reference wavelength of \mbox{$\lambda_0$ = 300\,$\mu$m)} are often chosen. This approach and these specific values can be traced back to \citet{Hildebrand1983} where they were derived for the reflection nebula NGC\,7023. However, \citet{Hildebrand1983} states, that the value for $\kappa_{\nu_0}$ is "probably good within a factor of 3 or 4 for measured cloud NGC\,7023". In a few other cases, e.g., \citet{Hill2009, Fissel2016}, slightly different values were adopted: $\kappa_0$ = 1.0 cm$^2$ g$^{-1}$, $\lambda_0$ = 1.2\,mm (corresponding to a reference frequency of  $\nu_0$ = 250\,GHz).  We explicitly note, that although these values are widely used, it is not obvious that they are indeed applicable for different objects. 
Alternatively Sil-Graph is often used in other studies. Whether this dust composition is indeed reflecting the actual dust properties for a given source better than the power-law approach is not obvious.
As it will be shown subsequently, the inconclusiveness regarding the dust properties introduces the greatest uncertainty of the results derived with this method. \\
A comparison between the dust opacities for the two different approaches is shown in Fig. \ref{Figure:Overview_Dust_Properties}.
The slope is similar in all cases, but the absolute deviations are about one order of magnitude in the most extreme case. Due to these deviations, the results from the fitting process are different as well (see Fig. \ref{Figure:Fit_NH2_Beta_T_5e2Msun_T_const20K_all_wl}).
Since the slope of the different dust opacities is similar, the fitting results for the column density $N$(H$_2$) show similar slopes as well, but the absolute values show deviations of up to $\sim$ 80\,\%. 
For a more comprehensive analysis of the impact of different dust models on the fitting results see Sect. \ref{Section:Different_Dust_Models}. The fitting results displayed in Fig. \ref{Figure:Fit_NH2_Beta_T_5e2Msun_T_const20K_all_wl} also illustrate that the derived values using the BFM are highly accurate if the dust opacities are well constrained. The very small deviations at the outer edge of the object are due to beam convolution, since the discrete outer edge is "smeared out" due to the convolution.
In addition to the column density $N$(H$_2$) and temperature $T$, the dust emissivity index $\beta$ is part of the fitting results as well, if the power-law approach with $\beta$ as a free parameter is adopted. In this case a constant dust emissivity index $\beta$ = 1.94 is derived. A constant value for $\beta$ is expected since the dust properties are constant within our model. 
As the possible variability of the dust properties along a given line of sight (e.g., due to freeze-out of molecules in the densest regions of  molecular cloud cores), $\beta$ would be spatially variable as well. However, this additional complexity is outside the focus of the current study.
A more detailed discussion on $\beta$ is provided in Sect. \ref{Section:Different_Dust_Models}. \\ 
Since the power-law approach shows rather high deviations (\mbox{$\sim$ 80\,\%} for the column density), which could potentially conceal additional uncertainties in the method, we focus on the impact of selected parameters on the fitting results for the ideal case in the following sections.

\subsection{Effect of different observing wavelengths on the fitting results} \label{Section:Fitting_procedure_for_selected_wavelengths}

\begin{figure*} 
\includegraphics[width=\textwidth]{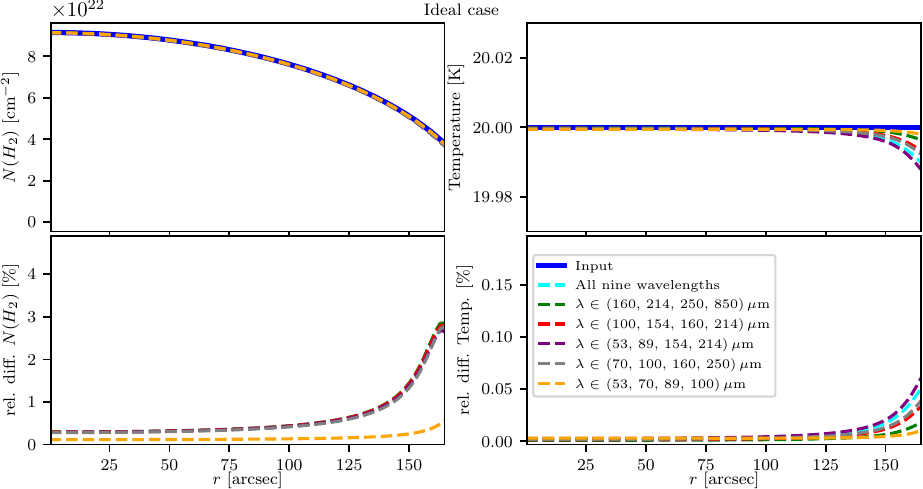} \caption{\: Results of the fitting process in the ideal case for the reference model when using different selections of observing wavelengths. \textit{Top:} Radial profiles of column density and temperature. The blue line marks the input values for the column density and temperature, respectively. \textit{Bottom:} Relative deviations between the results of the fitting process and the input values for column density (left) and temperature (right). Adopting the wavelength combination (54, 70, 89, 100)$\,\mu$m yields slightly lower deviations for the column density in comparison to the other wavelength selections, which yield very similar deviations.} \label{Figure:Overview_Results_Fitting_Process_Benchmark_Model_Selective_wl}
\end{figure*}

Often there are fewer observations than at the assumed nine wavelengths (Sect. \ref{Section:Model}) available. In this section we analyze the impact of a partial wavelength coverage on the derived parameters.
The following wavelength coverages are considered: a) all wavelengths, b) only 53, 89, 154, 214\,$\mu$m (HAWC+), c) only 70, 100, 160, 250\,$\mu$m (Herschel), d) only the four shortest wavelengths (53, 70, 89, 100$\,\mu$m), e) only the four longest wavelengths (160, 214, 250, 850$\,\mu$m), and f) only four intermediate wavelengths (100, 154, 160, 214$\,\mu$m). The corresponding intensity maps are beam-convolved to the lowest resolution available in each selection. \\
In Fig. \ref{Figure:Overview_Results_Fitting_Process_Benchmark_Model_Selective_wl} the results obtained with the fitting routine for the different selections are shown. 
\begin{figure*} 
\includegraphics[width=\textwidth]{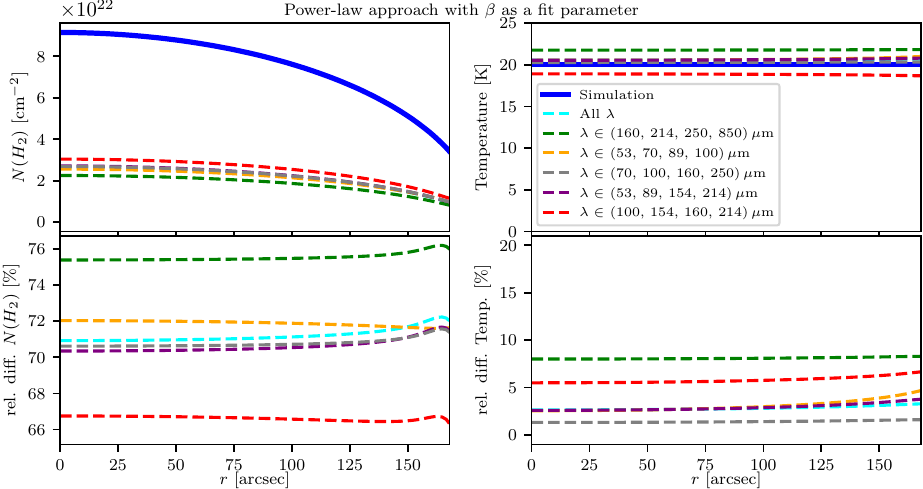} 
\caption{\: Same as Fig. \ref{Figure:Overview_Results_Fitting_Process_Benchmark_Model_Selective_wl}, but the power-law approach with $\beta$ as a fitting parameter was used to determine the dust properties. } \label{Figure:Overview_Results_Fitting_Process_Benchmark_Model_Selective_wl_power_law_b_free}
\end{figure*}
If the dust properties are known, the resulting column density and temperature for each selection of observing wavelengths show deviations smaller than 5\%.
The results are similar with the exception of those obtained for the selection of observing wavelengths of (53, 70, 89, 100)\,$\mu$m.
The reason for the deviation in the case of the selection of the shortest observing wavelengths is higher angular resolution (beam size is 6.8$''$ compared to 18.1/18.7$''$ for the other wavelength selections, see Tab. \ref{Table:Overview_instruments_wavelengths}). Measurements at any four out of the nine selected wavelengths yield sufficiently accurate results as long as the dust properties are known. \\
Using the power-law approach (for all $\beta$ values) for the dust properties the conclusions are similar: the results of the fitting process for each selection of observing wavelengths are comparable to these results obtained on the basis of measurements at all wavelengths (deviations $<$ 10\%; see Fig. \ref{Figure:Overview_Results_Fitting_Process_Benchmark_Model_Selective_wl_power_law_b_free}).
While the qualitative slope of the column density is comparable, there are still differences of about 10\,\%. This behaviour can be explained by the fact that, in addition to column density and temperature, the dust emissivity index $\beta$ is also a fit parameter (see Tab. \ref{Table:Beta_Values_Deviation_NH2} for a list of values of $\beta$ and corresponding deviation for each selection of observing wavelengths).
\begin{table*} 
  \begin{center}
    \caption{\: Dust emissivity index $\beta$ and relative differences between the results of the fitting process and the input values for column density for different selections of observing wavelengths.}
    \label{Table:Beta_Values_Deviation_NH2}
  \noindent   \begin{tabular}{cccc}
    \hline \hline    \rule{0pt}{3ex}
     Observing wavelengths & $\beta$ & rel. diff. $N$(H$_2$) [\%]  \\
      \hline \hline
      \rule{0pt}{3ex}
     \noindent (160, 214, 250, 850)\,$\mu$m & 1.86 & 75.5   \\ \rule{0pt}{2ex}
     \noindent (53, 70, 89, 100)\,$\mu$m & 1.90 & 71.9  \\ \rule{0pt}{2ex}  
     \noindent (53, 89, 154, 214)\,$\mu$m & 1.91 & 70.6   \\ \rule{0pt}{2ex}     
     \noindent All $\lambda$ & 1.94 & 71.2   \\ \rule{0pt}{2ex}  
     \noindent (70, 100, 160, 250)\,$\mu$m & 2.03 & 70.8   \\ \rule{0pt}{2ex}
     \noindent (100, 154, 160, 214)\,$\mu$m & 2.44 & 66.6   
\\      \hline \hline
    \end{tabular} 
  \end{center}
\end{table*}
While the choice of observing wavelengths has little effect on the resulting values of the column density and temperature (deviations between the different selections are $<$ 10\,\%), it has a higher impact on the dust emissivity index. See Sect. \ref{Section_Influence_Dust_Properties} for more details about $\beta$.

\subsection{Relevance of ALMA observations} \label{Section:ALMA_wavelengths}
\begin{figure*} [h!]
\includegraphics[width=\textwidth]{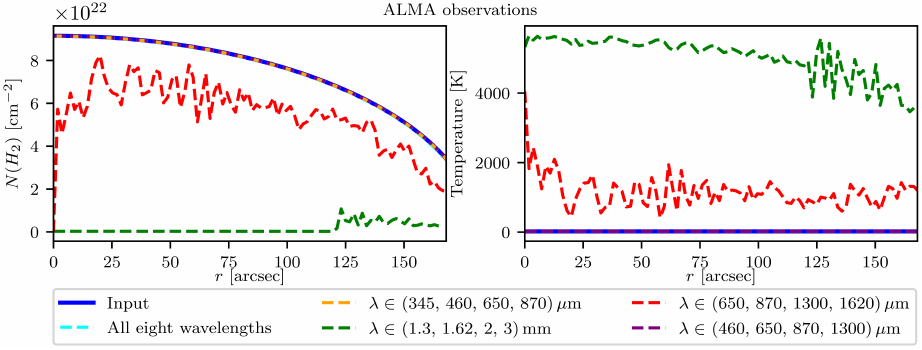} \caption{\: Results of the fitting process for the column density and temperature in the ideal case for the reference model when using different selections of ALMA observing wavelengths. With the exceptions of (650, 870, 1300, 1620)$\,\mu$m and (1.3, 1.62, 2, 3)\,mm, all ALMA wavelength selections yield very similar results.} \label{Figure:Overview_Results_ALMA_Wavelengths}
\end{figure*}

With the end of observations with Herschel Space Observatory \citep{Pilbratt2010} in 2013 and SOFIA \citep{Temi2018} in 2022, no observatory is currently available that covers the \mbox{$\sim$ 30-250\,$\mu$m}\footnote{Balloon-borne experiments such as the Balloon-borne Large Aperture Submillimeter Telescope for Polarimetry  \citep[BLASTPol,][]{Galitzki2014} and the Polarized Instrument for Long wavelength Observations of the Tenuous interstellar medium \citep[PILOT,][]{Bernard2016} allow one observations at a wavelength of 250\,$\mu$m. However, the angular resolution that can be achieved with them is lower than that of the observatories considered in our study.} wavelength range. For this reason, we now investigate the potential of the BFM if observations with the Atacama Large Millimeter/submillimeter Array (ALMA) are considered.
We consider ideal observations at eight wavelengths stated in the ALMA technical handbook\footnote{\url{https://almascience.nrao.edu/documents-and-tools/cycle9/alma-technical-handbook}, Table 7.1.} for cycle 9 (see Tab. \ref{Table:Possible_ALMA_wavelengths}). Regarding the resolution, we adopt the values for the 7-m array configuration. As in Sect. \ref{Section:Fitting_procedure_for_selected_wavelengths}, we perform the fitting using measurements at different combinations of wavelengths: all eight wavelengths, only the four shortest wavelengths (345/460/650/870$\,\mu$m), only the four longest wavelengths (1.3/1.62/2/3$\,$mm), and two intermediate wavelength ranges (650/870/1300/1620\,$\mu$m; 460/650/870/1300\,$\mu$m). Here we present the fitting results using the exact dust properties. As before, the corresponding intensity maps are beam-convolved to the lowest resolution in each selection of wavelengths (see Fig. \ref{Figure:Overview_Results_ALMA_Wavelengths} for the result of the fitting process using the different observing wavelength combinations). \\
When considering measurements either at all wavelengths, at the four shortest wavelengths or at \\ 460/650/870/1300\,$\mu$m, the fitting yields accurate results. 
When considering observations at the four longest wavelengths and at the combination 650/870/1300/1620\,$\mu$m, the derived values deviate by multiple orders of magnitude from the input values. 
Our fitting results show that at least one observing wavelength needs to be $\le$ 460\,$\mu$m to constrain the location of the
maximum of the SED for a spherical dust cloud with a temperature of 20\,K. If dust distributions with a higher temperature were to be considered, shorter observing wavelengths would be required.
 However, observations at ALMA band 9 and 10 are challenging mainly due to the difficulty of finding sufficiently bright calibrators and low atmospheric transmission \citep{Cortes2021}. \\
In conclusion, based on observational wavelengths that can constrain the wavelength maximum of the SED, the fitting results of the BFM adequately represent the column density and temperature.
If the selection of observing wavelengths allows only a weak constraint on the temperature, then the results of the fit will be unreliable due to the degeneracy between column density and temperature.

\begin{table*} 
  \begin{center}
    \caption{\: Overview of possible observing wavelengths in the case of ALMA cycle 9 (7-m array). For more information see ALMA Cycle 9 Technical Handbook \citep[Table 7.1;][]{Cortes2021}.}
    \label{Table:Possible_ALMA_wavelengths}
  \noindent   \begin{tabular}{ccccc}
    \hline \hline    \rule{0pt}{3ex}
      Band & Wavelength [$\mu$m] & Beam size (7-m) [$''$]  & Uncertainty [\%]  \\
      \hline \hline
      \rule{0pt}{3ex}
     \noindent 3 & 3000 & 12.5  & 10 \\ \rule{0pt}{2ex}
     \noindent 4 & 2000 & 8.35  & 10 \\ \rule{0pt}{2ex}  
     \noindent 5 & 1620 & 6.77  & 10 \\ \rule{0pt}{2ex}     
     \noindent 6 & 1300 & 5.45  & 10 \\ \rule{0pt}{2ex}
     \noindent 7 & 870 & 3.63  & 10 \\ \rule{0pt}{2ex}     
     \noindent 8 & 650 & 2.72  & 10 \\ \rule{0pt}{2ex}     
     \noindent 9 & 460 & 1.83  & 10 \\ \rule{0pt}{2ex}     
     \noindent 10 & 345 & 1.44  & 10 
\\      \hline \hline
    \end{tabular} 
  \end{center}
\end{table*}

\section{Results II: Impact of the characteristics of the source} \label{Section:Characteristics_of_the_source}
In the following, we analyze the impact of the source characteristics on the reliability of the BFM.

\subsection{Influence of the optical depth $\tau$} \label{Section:Influence_of_the_optical_depth}
While the assumption of optically thin emission in the considered wavelength range is valid for most molecular clouds, it may break down for the densest region and cores. Therefore, in the following we investigate to what extent the BFM provides reliable results at higher optical depths, i.e., beyond the optically thin case represented by our reference model (Sect. \ref{Section:Model}). 
We study gas masses up to multiple orders of magnitude higher than in the reference model so that the optical depth at the highest observing wavelength is $\tau_{850\,\mu\text{m}}$ $\gg$ 1.
In Tab. \ref{Table:List_optical_depths} we list the selected gas masses, the correlated maximum column densities and values for the corresponding maximum optical depth at the shortest (53\,$\mu$m) and longest (850\,$\mu$m) observing wavelengths (see Tab. \ref{Table:Overview_instruments_wavelengths}) and the calculated relative deviations.

\begin{figure*} 
\includegraphics[width=\textwidth]{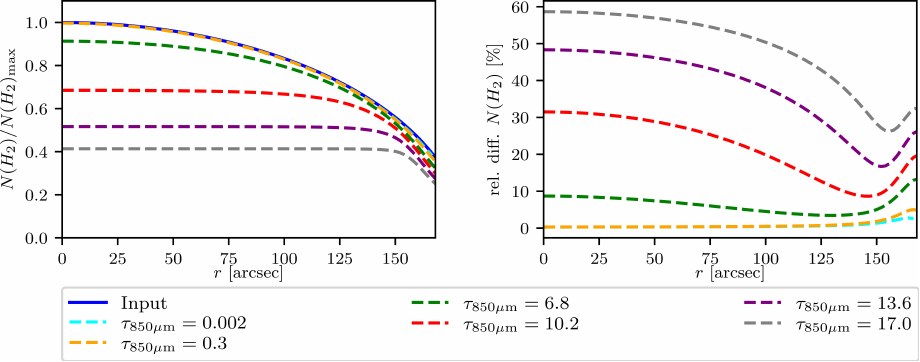} \caption{\: Overview of the fitting results in the ideal case for higher optical depths. The optical depth values stated here represent the maximum values at 850\,$\mu$m, i.e., at the core of the sphere. \textit{Left:} Radial profiles of column density. Since the column density covers several orders of magnitude, the profiles are normalized to the respective maximum input value.
 The solid lines mark the input values and the dashed lines the derived values for the column density. \textit{Right:} Relative deviations between the results of the fitting process and the input values.} \label{Figure:Fit_Results_Analyze_Higher_Optical_Depth}
\end{figure*}

\begin{table*} 
  \begin{center}
    \caption{\: Overview of optical depths $\tau_{\lambda}$ at the shortest (53\,$\mu$m) and longest (850\,$\mu$m) observing wavelengths we consider for the fitting process for different gas masses. Additionally, values for the maximum input column densities and the mean deviations for the column density ($\overline{d_{N\text{(H}_2)} }$) in the ideal case are listed.}
    \label{Table:List_optical_depths}
  \noindent   \begin{tabular}{cccccc}
    \hline \hline    \rule{0pt}{3ex}
     $M_\text{gas}$ [$M_\odot$] & $N$(H$_2$)$_\text{max}$ [cm$^{-2}$] & $\tau_{53\,\mu\text{m}}$  & $\tau_{850\,\mu\text{m}}$ & $\overline{d_{N{\text{(H}_2)}} }$ (ideal case) \\
      \hline \hline
      \rule{0pt}{3ex}
     \noindent  5$\cdot$10$^2$ & 9.15$\cdot$10$^{22}$  & 0.6  & 0.002 & 0.6\,\%  \\ \rule{0pt}{2ex}
     \noindent  1$\cdot$10$^3$ & 1.83$\cdot$10$^{23}$ & 1.2  & 0.003   &  0.6\,\% \\ \rule{0pt}{2ex}
     \noindent  1$\cdot$10$^4$ & 1.83$\cdot$10$^{24}$ & 11.5  &  0.03  & 0.7\,\% \\ \rule{0pt}{2ex}
     \noindent  1$\cdot$10$^5$ & 1.83$\cdot$10$^{25}$ & 115.0  &  0.3 & 0.9\,\%  \\ \rule{0pt}{2ex}
     \noindent  1$\cdot$10$^6$ & 1.83$\cdot$10$^{26}$ & 1150.4  &  3.4 &  2.2\,\%  \\ \rule{0pt}{2ex}
     \noindent  2$\cdot$10$^6$ & 3.66$\cdot$10$^{26}$ & 2300.8  &  6.8 & 6.8\,\%  \\ \rule{0pt}{2ex}
     \noindent  3$\cdot$10$^6$ & 5.50$\cdot$10$^{26}$ & 3451.2  &  10.2 & 21.6\,\%  \\ \rule{0pt}{2ex}
     \noindent  4$\cdot$10$^6$ & 7.32$\cdot$10$^{26}$ & 4601.6  &  13.6 & 36.4\,\%  \\ \rule{0pt}{2ex}
     \noindent  5$\cdot$10$^6$ & 9.15$\cdot$10$^{26}$ & 5752.0  & 17.0 &  47.3\,\% 
\\      \hline \hline
    \end{tabular} 
  \end{center}
\end{table*}

In all presented cases the temperature is accurately reconstructed via the BFM. When observations are tracing regions with higher optical depths, the fit delivers unreliable results up to 50\% deviation in the worst case in the parameter space considered here. More specifically, starting at an optical depth of \mbox{$\tau_{850\,\mu\text{m}}$  > 10}, the method underestimates the actual column density.
As the optical depth measured along the line of sight decreases, the slope of the deviations decreases with increasing distance from the centre. In the outermost region, the deviations increase as beam effects at the outer edge come into play, as already seen in the optically thin case (see Fig. \ref{Figure:Fit_NH2_Beta_T_5e2Msun_T_const20K_all_wl}).
 As long as observations are tracing regions with an optical depth \mbox{$\tau_{850\,\mu\text{m}}$  < 5}, the method delivers accurate results.
Since our model cloud is optically thin below column densities of $\sim$ 10$^{24}$ cm$^{-2}$, i.e., a value ranging at the upper boundary in the case of typical molecular clouds, the analysis applying the BFM is well justified. 
However, beyond this value the exponentially increasing attenuation of the emitted radiation with linearly decreasing column density, makes the BFM hardly applicable.
This is potentially important in the case of the densest, e.g., filamentary substructures of molecular cloud cores or even more in the case of embedded protoplanetary disks.
One example is the filamentary molecular cloud OMC-3. While this object is optically thin at FIR wavelengths on large scales \citep[\mbox{$\sim$ 10$^4$ au}; e.g.,][]{Zielinski2022}, \citet{Liu2021c} finds that the innermost region on a scale of $\sim$ 10$^2$ au around MMS6 observed with ALMA is optically thick even at 1.2\,mm.

\subsection{Influence of different dust models on the fitting results} \label{Section:Different_Dust_Models}
The BFM is often used to study (potentially) star-forming molecular clouds or filaments. Prior to the later evolution of the dust phase in protoplanetary disks to planetary cores and terrestrial planets, observations indicate dust grain growth to $\gtrsim$ 10$\,\mu$m sized grains already in class 0 envelopes \citep[e.g.,][]{Valdivia2019, Hull2020, LeGouellec2020}.

\begin{figure*} 
\includegraphics[width=\textwidth]{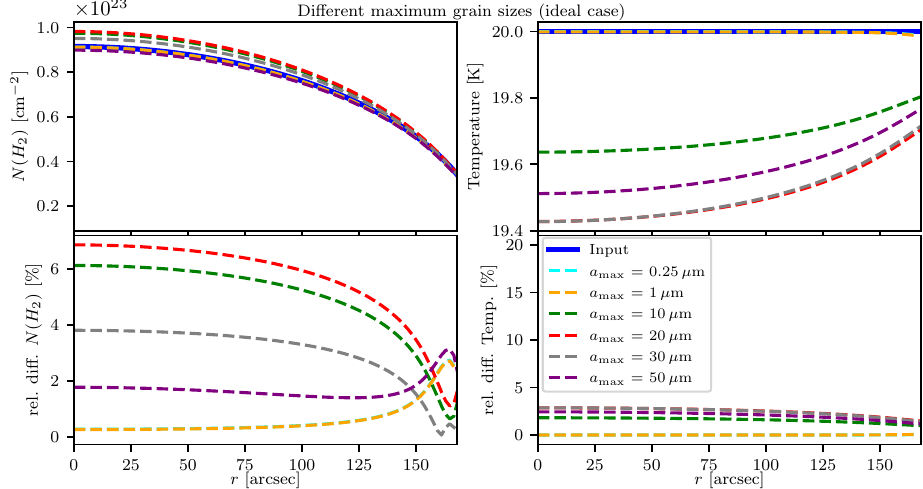} \caption{\: Results of the fitting process in the ideal case for the reference model when considering different maximum dust grain sizes. \textit{Top:} Radial profiles of column density and temperature. The blue line marks the input values for the column density and temperature, respectively. \textit{Bottom:} Relative deviations between the results of the fitting process and the input values for column density (left) and temperature (right). When considering $a_\text{max}$ = 0.25 and $a_\text{max}$ = 1\,$\mu$m, the results for the temperature are very similar.} \label{Figure:Overview_Fit_Col_Dens_T_higher_dust_sizes_ideal_case}
\end{figure*}

 In the following, we therefore analyze the impact of different dust grain sizes and compositions on the results derived with the BFM, while we focus on the ideal case and the power-law approach for the dust properties with $\beta$ as the fitting parameter.
We consider different maximum grain sizes $a_\text{max} \in$ (0.25, 1, 10, 20, 30, 50)\,$\mu$m, and different chemical compositions (Sil-Graph, pure astrosilicate, pure graphite). \\
The results for the ideal case are compiled in Tab. \ref{Table:Overview_relative_deviations_col_dens_temp_ideal_case}, while those for the power-law approach for the dust properties with $\beta$ as a fitting parameter are listed in Tab. \ref{Table:Overview_relative_deviations_col_dens_temp_power_law_case_b_free}.
In Fig. \ref{Figure:Overview_Fit_Col_Dens_T_higher_dust_sizes_ideal_case}, the BFM results for the reference model and Sil-Graph are shown for different dust grain sizes. The relative deviations of the derived temperature and column density are smaller than 5\,\% and 10\,\%, respectively. Interestingly, depending on the maximum grain size, the column density is under- or overestimated. 
This deviation can be traced back to the assumed simplification that the wavelength-dependent opacity $\kappa_\nu$ does not depend on the grain size.
For the reference model with dust grain sizes from 5\,nm to 250\,nm, i.e., well below the considered observing wavelengths, this assumption does not effect the results of the BFM. 
However, for significantly larger grains, the emission characteristics of the broader grain size distribution are not well represented in this approach.

We find similar relative deviations of the derived temperature and column density for different chemical compositions of the dust.
Considering graphite the BFM yields slightly higher deviations for $T$ and $N$(H$_2$) than in the case of Sil-Graph and silicate.

\begin{figure*} 
\includegraphics[width=\textwidth]{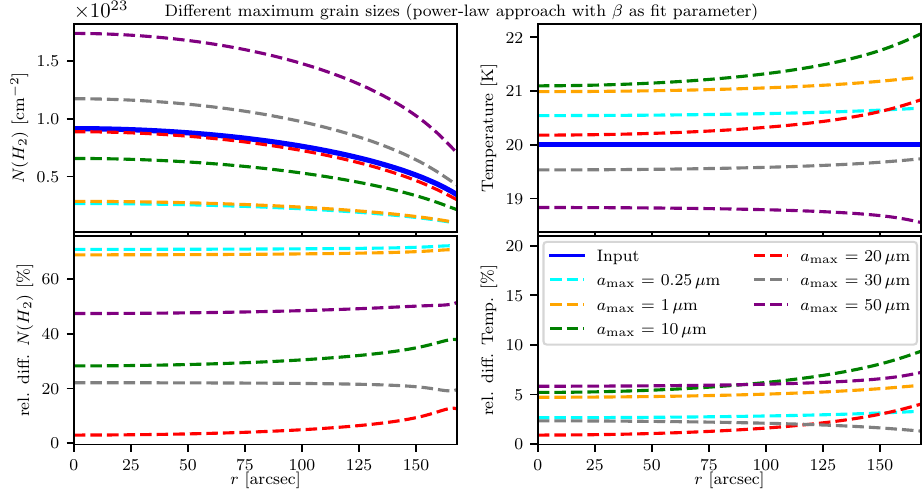} 
\caption{\: Same as Fig. \ref{Figure:Overview_Fit_Col_Dens_T_higher_dust_sizes_ideal_case}, but the power-law approach with $\beta$ as a fitting parameter was used to determine the dust properties. When considering $a_\text{max}$ = 0.25 and $a_\text{max}$ = 1\,$\mu$m, the results for the column density are very similar.} \label{Figure:Overview_Fit_Col_Dens_T_higher_dust_sizes_pow_case_b_free}
\end{figure*}

 Fig. \ref{Figure:Overview_Fit_Col_Dens_T_higher_dust_sizes_pow_case_b_free} shows the fitting results for the reference model using a power-law approach for the dust properties and $\beta$ as a fitting parameter.
The relative deviations with respect to the temperature are up to 10\,\% and the relative deviations with respect to the column density are up to 70\,\%.  Depending on the dust size and composition, the column density is vastly under- or overdetermined, between +50\,\% and -70\,\%. The BFM results considering different power-law approaches for the dust properties ($\beta$ = 1.62, 2) are qualitatively similar. 

\begin{table*} 
  \begin{center}
    \caption{\: Overview of relative deviations for column density and temperature in the ideal case for different maximum dust grain sizes and chemical compositions. \textcolor{teal}{Positive} (\textcolor{red}{negative}) values indicate that the considered quantity ($N$(H$_2$), $T$) has been overestimated (underestimated).}
    \label{Table:Overview_relative_deviations_col_dens_temp_ideal_case}
  \noindent   \begin{tabular}{ccccccc}
    \hline \hline    \rule{0pt}{3ex}
     & $\overline{d_{N{\text{(H}_2}), \text{ideal}} }$ [\%] & $\overline{d_{N\text{(H}_2), \text{ideal}} }$ [\%] & $\overline{d_{N\text{(H}_2), \text{ideal}} }$ [\%] & $\overline{d_{T, \text{ideal}} }$ [\%] & $\overline{d_{T, \text{ideal}} }$ [\%] & $\overline{d_{T, \text{ideal}} }$ [\%] \\
     $a_\text{max}$ [$\mu$m] & Silicate & Graphite  & Sil-Graph & Silicate & Graphite  & Sil-Graph \\
      \hline \hline
      \rule{0pt}{3ex}
     \noindent 0.25  & \textcolor{red}{ -0.6} & \textcolor{red}{ -0.6} & \textcolor{red}{ -0.6} & 0.0 & 0.0 &  0.0  \\ \rule{0pt}{2ex}
     \noindent 1  & \textcolor{red}{ -0.6}  & \textcolor{red}{ -0.6} & \textcolor{red}{ -0.6} & 0.0 & 0.0 &  0.0  \\ \rule{0pt}{2ex}
     \noindent 10  & \textcolor{teal}{+3.5}  & \textcolor{teal}{+7.0} & \textcolor{teal}{+4.9} &  \textcolor{red}{ -1.2} & \textcolor{red}{-2.2} & \textcolor{red}{ -1.6}  \\ \rule{0pt}{2ex}
     \noindent 20  & \textcolor{teal}{+3.9}  & \textcolor{teal}{+8.3} & \textcolor{teal}{+5.6} &  \textcolor{red}{ -1.8}  & \textcolor{red}{ -3.5} & \textcolor{red}{ -2.5}  \\ \rule{0pt}{2ex}
     \noindent 30  & \textcolor{teal}{+1.6}  & \textcolor{teal}{+5.3} & \textcolor{teal}{+3.0} & \textcolor{red}{ -1.8} & \textcolor{red}{-3.5} & \textcolor{red}{-2.5} \\ \rule{0pt}{2ex}
     \noindent 50  &  \textcolor{red}{-2.8} & \textcolor{teal}{+0.4} & \textcolor{red}{-1.7} &   \textcolor{red}{-1.4} & \textcolor{red}{-2.4} & \textcolor{red}{-3.1}  
\\      \hline \hline
    \end{tabular} 
  \end{center}
\end{table*}

\begin{table*}
  \begin{center}
   \caption{\: Same as Tab. \ref{Table:Overview_relative_deviations_col_dens_temp_ideal_case}, but the power-law approach with $\beta$ as a fitting parameter was used to determine the dust properties. } \label{Table:Overview_relative_deviations_col_dens_temp_power_law_case_b_free}
  \noindent   \begin{tabular}{ccccccc}
    \hline \hline    \rule{0pt}{3ex}
     & $\overline{d_{N\text{(H}_2), \: \beta = \text{free}} }$ [\%] & $\overline{d_{N\text{(H}_2), \: \beta = \text{free}} }$ [\%] & $\overline{d_{N\text{(H}_2), \: \beta = \text{free}} }$ [\%] & $\overline{d_{T, \: \beta = \text{free}} }$ [\%] & $\overline{d_{T, \: \beta = \text{free}} }$ [\%] & $\overline{d_{T, \: \beta = \text{free}} }$ [\%] \\
     $a_\text{max}$ [$\mu$m] & Silicate & Graphite  & Sil-Graph & Silicate & Graphite  & Sil-Graph \\
      \hline \hline
      \rule{0pt}{3ex}
     \noindent 0.25  & \textcolor{red}{-68.6} & \textcolor{red}{-74.9} & \textcolor{red}{-71.2} & \textcolor{teal}{+1.6} & \textcolor{teal}{+4.4} &  \textcolor{teal}{+2.8}  \\ \rule{0pt}{2ex}
     \noindent 1  & \textcolor{red}{-68.7}  & \textcolor{red}{-69.3} & \textcolor{red}{-69.3} & \textcolor{teal}{+1.7} & \textcolor{teal}{+8.8} &  \textcolor{teal}{+5.0}  \\ \rule{0pt}{2ex}
     \noindent 10  & \textcolor{red}{-68.9}  & \textcolor{teal}{+31.2} & \textcolor{red}{-30.8} & \textcolor{teal}{+3.9} & \textcolor{teal}{+2.9} & \textcolor{teal}{+6.3}  \\ \rule{0pt}{2ex}
     \noindent 20  & \textcolor{red}{-54.3}  & \textcolor{teal}{+46.9} & \textcolor{red}{-5.3} &  \textcolor{red}{-2.4}  & \textcolor{red}{-1.0} & \textcolor{teal}{+1.7}  \\ \rule{0pt}{2ex}
     \noindent 30  & \textcolor{red}{-24.9}  & \textcolor{teal}{+49.2} & \textcolor{teal}{+21.6} & \textcolor{red}{-5.9} & \textcolor{red}{-1.2} & \textcolor{red}{-2.1} \\ \rule{0pt}{2ex}
     \noindent 50  &  \textcolor{teal}{+29.0} & \textcolor{teal}{+33.1} & \textcolor{teal}{+47.5} &  \textcolor{red}{-10.0} & \textcolor{red}{-0.8} & \textcolor{red}{-6.1}  
\\      \hline \hline
    \end{tabular} 
  \end{center}
\end{table*}

These results underline that uncertainties about the dust properties represent the greatest weakness of the BFM. 
In most studies in which this method is utilized the same dust model is assumed (see Sect. \ref{Section:Details_about_the_blackbody_fit_method_and_the_model_space}). While this approach provides a basis for a comparison of results obtained in individual studies, it is highly uncertain whether this assumption is indeed justified given the range of physical properties in the various types of astrophysical objects. 

\subsubsection*{Dust emissivity index $\beta$}
In the case of the power-law approach with $\beta$ as a fitting parameter, the dust emissivity index $\beta$ is -- in addition to the column density and the temperature (see e.g., Fig. \ref{Figure:Fit_NH2_Beta_T_5e2Msun_T_const20K_all_wl}) -- derived as well.
If the dust properties are known, the dust emissivity index cannot be constrained using the BFM, since Eq. \ref{Equation:Dust_Opacity} is not adopted. 
However, $\beta$ can be determined from the slope of the SED\footnote{In the following, we refer to $\beta_\mathrm{SED}$ as the dust emissivity index, which is calculated on the basis of the SED, and $\beta_\mathrm{fit}$ as the dust emissivity index resulting from the BFM.} \citep[e.g.,][]{Friesen2005}: 
\begin{equation} \label{Equation:Beta_SED}
\gamma = \frac{\log(S_2/S_1)}{\log(\nu_2/\nu_1)} = \beta_\mathrm{SED} + \alpha \: .
\end{equation}
In the case of the Rayleigh-Jeans approximation \citep{Rayleigh1900, Jeans1905}, $\alpha$ = 2 and thus $\beta_\mathrm{SED} = \gamma - 2$.
In order to apply the Rayleigh-Jeans approximation, $\frac{h\nu}{kT} \ll 1$ must hold.
Based on our selection of observing wavelengths (see Tab. \ref{Table:Overview_instruments_wavelengths}), this approximation cannot be applied since $T \gg 58$\,K must be satisfied for $\lambda$=250\,$\mu$m and $T \gg 17$\,K for $\lambda$=850\,$\mu$m. 
This constraint is in contradiction to the constant dust temperature in our reference model of 20\,K.
Therefore, we derive $\beta_\mathrm{SED}$ based on additional simulated measurements at \mbox{$\lambda$ = 1000\,$\mu$m} and \mbox{$\lambda$ = 2000\,$\mu$m}\footnote{We note that these wavelengths are only used for the calculation of $\beta_\text{SED}$ and not for the fitting process itself.} for which $T\gg14\,$K and $T \gg 7\,$K are satisfied, respectively. \\

\begin{figure*} 
\includegraphics[width=\hsize]{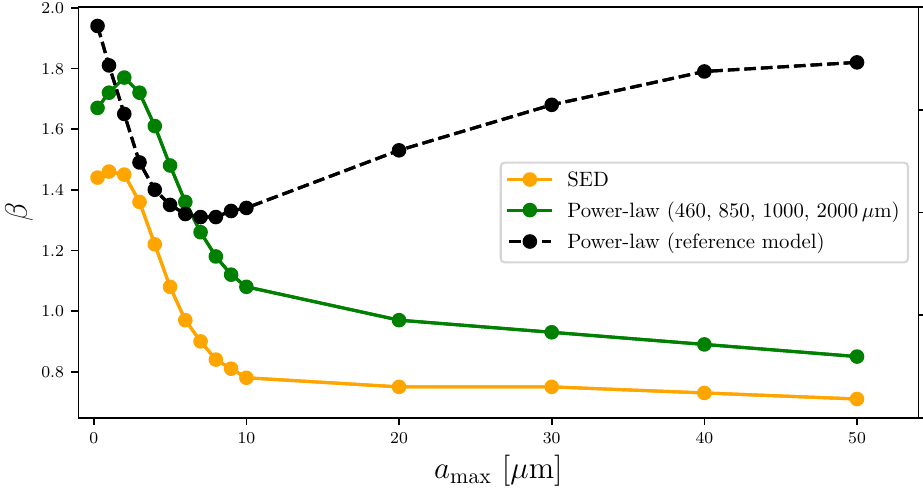} \caption{\: Dust emissivity index $\beta$ as a function of the maximum dust grain size $a_\mathrm{max}$. Power-law: $\beta_\mathrm{fit}$ derived using Eq. \ref{Equation:Dust_Opacity}; SED: $\beta_\mathrm{SED}$ determined using Eq. \ref{Equation:Beta_SED}.} \label{Figure:Overview_Beta_vs_amax}
\end{figure*}

While a dust emissivity index of $\beta$ $\lesssim$ 2 is found for the interstellar medium \citep{PlanckCollaboration2011B}, lower values are derived for star-forming regions \citep[$\approx $ 2--1.5; e.g.,][]{Shirley2005, Beuther2007, Sadavoy2013} and protoplanetary disks \citep[$<$1; e.g.,][]{Isella2010, Tripathi2018, Galametz2019}. Given the relation between the wavelength-dependent absorption and thus emission efficiency and grain size, large $\beta$-values indicate the dominance of small grains, while small $\beta$-values that of larger grains in the contribution to net emission in the FIR to mm wavelength range \citep[][]{Natta2004b, Draine2006}.  
In this section, larger dust grains are considered to evaluate the consistency of these two different methods to derive $\beta$.\\
In Fig. \ref{Figure:Overview_Beta_vs_amax}, the dust emissivity index $\beta$ -- derived using the two methods mentioned above -- is plotted as a function of the maximum dust grain size $a_\mathrm{max}$.
The result obtained via the BFM ($\beta_\mathrm{fit}$, Eq. \ref{Equation:Dust_Opacity}) is marked with "power-law", while the result using the slope of the SED in the Rayleigh-Jeans regime ($\beta_\mathrm{SED}$, Eq. \ref{Equation:Beta_SED}) is marked with "SED". 
With regard to "power law", we distinguish two cases: Results for the reference model (see Table \ref{Table:Overview_instruments_wavelengths}) and results based on longest possible observing wavelengths which contain the auxiliary observing wavelengths \mbox{$\lambda$ = 1000\,$\mu$m} and \mbox{$\lambda$ = 2000\,$\mu$m.} \\
Apart from a minor increase ($<$1\,\%) at grain sizes between 1 and 2$\,\mu$m, the dust emissivity index $\beta_\mathrm{SED}$ is a strictly monotonically decreasing function of the dust grain size.
For small dust grains\footnote{We note that in general, sub-micrometer sized dust grains are expected in the ISM and molecular clouds, respectively \citep[e.g.,][]{MathisRumplNordsieck1977}. In this context, we use "small" or "larger" dust grains to refer to our selected dust grain radius of 0.25$\,\mu$m--50$\,\mu$m.} ($a_\mathrm{max}$ $\lesssim$ 7\,$\mu$m), $\beta_\mathrm{fit}$ also decreases with increasing maximum dust grain size. This finding supports the assumption that $\beta$ can serve as an indicator for the dust size and thus allows deriving constraints on the process of grain growth. However, $\beta_\mathrm{fit}$ increases significantly for larger dust grains (\mbox{$a_\mathrm{max}$ $\gtrsim$ 10\,$\mu$m}).
Considering the dust emissivity index derived based on longer observing wavelengths, $\beta_\mathrm{fit, (460, 850, 1000, 2000\,\mu\mathrm{m})}$ shows similar behaviour as $\beta_\mathrm{SED}$: Apart from an increase between 0.25 and 2\,$\mu$m, $\beta_\mathrm{fit, (460, 850, 1000, 2000\,\mu\mathrm{m})}$ is a monotonically decreasing function of dust grain size\footnote{We explicitly point out that this does not imply that $\beta$ should be derived with fewer observing wavelengths in general (see Sect. \ref{Section:Fitting_procedure_for_selected_wavelengths}), but that the range of observing wavelengths adopted is relevant.}.\\
The found qualitative mismatch between the values of $\beta$ derived with the different methods highlights a fundamental problem regarding $\beta_\mathrm{fit}$: On the one hand, the observing wavelengths should be sufficiently long to fulfill the Rayleigh-Jeans approximation, allowing $\beta_\mathrm{fit}$ to serve as indicator for the dust grain size. However, on the other hand, at least one short wavelength ($\le$ 460\,$\mu$m, see Sect. \ref{Section:ALMA_wavelengths}) needs to be part of the observing wavelength selection to constrain the location of the maximum of the SED in order to correctly derive $N$(H$_2$), $T$, and $\beta$ via the BFM. \\
In contrast to $\beta_\mathrm{SED}$, the quantity $\beta_\mathrm{fit}$ -- derived using the BFM -- is not suitable for making statements about the grain size.\\
Selected earlier studies indicated a potential correlation between dust temperature $T$ and dust emissivity index $\beta$ \citep[e.g.,][]{Dupac2001, Dupac2003, Chuss2019}. However, an analysis of this connection is not within the scope of this study (due to considering constant dust temperature $T$ and uniform dust properties, i.e., constant $\beta$, in our reference model) and we refer to \citet{Shetty2009a}, \citet{Shetty2009} and \citet{Anderson2012} for a more detailed discussion on this topic instead.

\subsubsection{Deriving the optical depth $\tau$} \label{Section:Fitting_optical_depth}
The results shown in Sect. \ref{Section:Different_Dust_Models} demonstrate that uncertainties about the dust properties represent the greatest weakness of the BFM.
Thus, we propose to first derive the optical depth $\tau$, and consider the column density in the second step in combination with a suitable dust model. If no constraints on the dust properties are available for the object of interest, considering different dust models \citep[e.g.,][]{MathisRumplNordsieck1977, Draine1984, Weingartner2001, Draine2009, Koehler2015, Draine2021} and their impact on the derived values of $N$(H$_2$) and $T$ appears mandatory. 
 
\begin{figure*} 
\includegraphics[width=\textwidth]{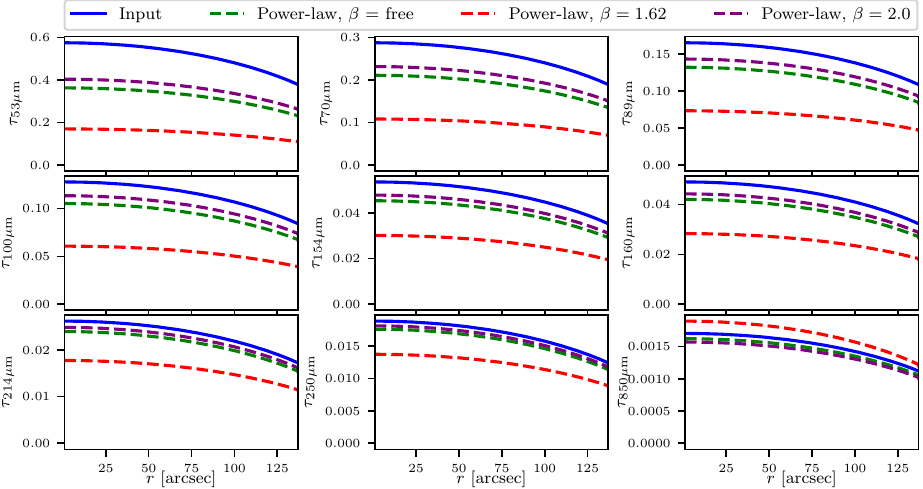} \caption{\: Results of the fitting process with respect to the optical depth $\tau_\lambda$ as a function of the distance to the center in the case of the reference model. Results are shown for the power-law approach with $\beta$ = (1.62, 2.0) and $\beta$ as a fitting parameter. When considering the power-law approach with $\beta$ = 2 and $\beta$ as a fitting parameter the results are similar, especially at higher optical depths.} \label{Figure:Results_Fit_Optical_Depth_all_nine_wl}
\end{figure*}

Without utilizing a specific dust model -- apart from the frequency dependence -- constraints can still be obtained about the object of interest, i.e., for the optical depth $\tau$, temperature $T$ and dust emissivity index $\beta$.
In the following, we present the results obtained with the BFM with respect to the optical depth $\tau$.
The specific fitting equation in this case is given by
\begin{equation} \label{Formula:Fit_Optical_Depth}
I_\nu = \left( 1 - \exp\left(- \epsilon \left( \nu/\nu_0 \right)^\beta \right) \right) \, \frac{2h\nu^3}{c^2} \, \frac{1}{\exp \left( \frac{h\nu}{kT} \right) - 1} \: .
\end{equation}
Here, the fitting parameters are $\epsilon$, $\beta$ and $T$. The optical depth at each respective observing wavelength is then calculated via Eq. \ref{Formular:Optical_depth}. The results for the optical depth at all nine considered observing wavelengths (see Tab. \ref{Table:Overview_instruments_wavelengths}) is given in Fig. \ref{Figure:Results_Fit_Optical_Depth_all_nine_wl}. \\
In the case of the power-law approach with $\beta$ as a fitting parameter, the relative differences for $\tau$ decrease with increasing wavelength, from $\sim$30-35\,\% (53$\,\mu$m) to <5\,\% (850$\,\mu$m). Considering observations at long wavelengths (>154$\,\mu$m), the BFM provides accurate results for the optical depth (relative deviation < 10\,\%), with the slope correctly derived in all cases. The same holds true for $\beta$ = 2. In the case of $\beta$ = 1.62, the deviations are higher (\mbox{70\,\% - <10\,\%}).
Considering other dust compositions (pure astrosilicate and pure graphite), the fitting results (for $\beta$=2 and $\beta$ as a fitting parameter) are comparable: In the case of astrosilicate the deviations are slightly lower (25 - <5\%) and in the case of graphite slightly higher (50 - 5\%). 
We assume that these deviations are caused by the approach to fit $\epsilon$, resulting in greater uncertainties at shorter wavelengths. The temperature is derived correctly in all cases, with relative deviations < 10$\,\%$.
As the choice of the value for $\epsilon$ can be traced back to the choice of the reference opacity $\kappa_{\nu, }$ and corresponding frequency $\nu_0$, we tested the BFM for values of $\nu_0$ $\in$ (100, 10$^6$)\,GHz and obtained the same results. 
 As long as the reference frequency $\nu_0$ is in approximately the same order of magnitude as the observing frequencies (see Tab. \ref{Table:Overview_instruments_wavelengths}), the results do not change. \\
In summary, the BFM provides reliable estimates of the optical depth as well as temperature, regardless of the assumed dust grain model (especially for measurements at longer wavelengths). 

\subsection{Embedded heating sources} \label{Section:Case_of_non_constant_temperature}
The BFM is often adopted for objects in which protostars have already formed. These stellar sources heat the surrounding dust, nullifying the assumption of a constant temperature. 
According to earlier studies, the BFM has been used to determine dust temperatures in the range of $\sim$ 20--100\,K \citep{Fissel2016, Chuss2019, Santos2019}. In the following, we study the impact of a non-constant temperature on the fit results.

\subsubsection*{Linear temperature distribution}

\begin{figure*}
\includegraphics[width=\textwidth]{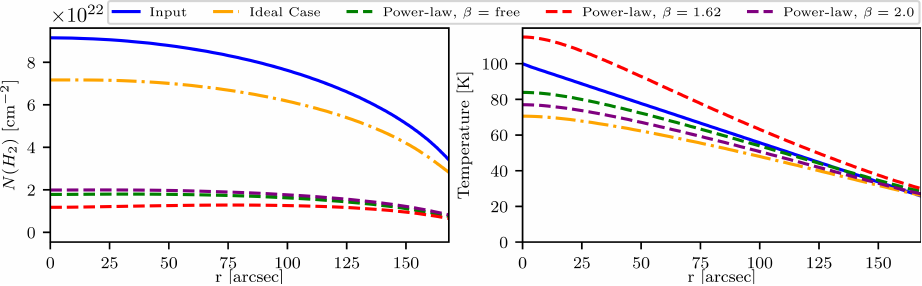} \caption{\: Results of the fitting process in the case of the reference model but with a temperature which is linearly decreasing from 100\,K (at the center) to 20\,K (at the outer edge). The blue line marks the
input values for the column density and spherical temperature distribution, respectively. When considering the power-law approach with $\beta$ = 2 and $\beta$ as a fitting parameter the results for the column density are similar.} \label{Figure:Benchmark_Model_T_var_100_10K}
\end{figure*}

We first consider the case of a linear temperature distribution with temperature decreasing from 100\,K at the center to 20\,K at the outer edge of the model space. The results obtained with the BFM are shown in Fig. \ref{Figure:Benchmark_Model_T_var_100_10K} for all cases described in Sect. \ref{Section:Description_of_the_fitting_process}, i.e., ideal case, power-law approach with $\beta$ $\in$ (2. 1.62) and $\beta$ as a fitting parameter. \\
The input temperature shown in Fig. \ref{Figure:Benchmark_Model_T_var_100_10K} represents the radial temperature distribution. Along the line of sight, the "effective" temperature is derived from the superposition of the radiation from regions with different local temperatures. Consequently, the temperature distribution along the line of sight can not be derived and thus a direct comparison with the temperature resulting from the BFM cannot be made.  
Therefore, the BFM can only be used to determine an approximate temperature distribution, potentially localizing embedded sources. This is possible because in those objects for which the BFM is applied regions with higher temperature are usually also denser, i.e., their relative contribution to the net flux/SED along the line of sight is higher than of the surrounding, colder environment (see Fig. \ref{Figure:Benchmark_Model_T_var_100_10K}).
The column density is derived with a deviation of $\lesssim$ 20\,\% in the ideal case.
However, the uncertainty regarding the dust properties (see Sect. \ref{Section_Influence_Dust_Properties}) is presumably much larger.

All power-law approaches for the dust properties yield similar deviations (up to 90\,\% for the column density). The temperature slope is correctly derived as well. 

\subsubsection*{Exponential temperature distribution}
\begin{figure*} 
\includegraphics[width=\textwidth]{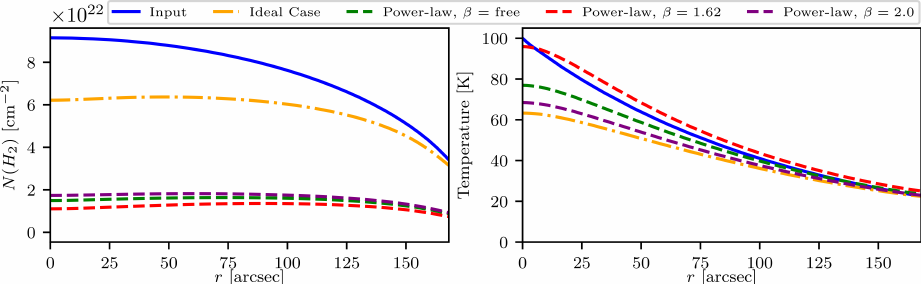} 
\caption{\: Same as Fig. \ref{Figure:Benchmark_Model_T_var_100_10K}, but with the temperature exponentially decreasing from 100\,K (at the center) to 20\,K (at outer edge). When considering the power-law approach with $\beta$ = 2 and $\beta$ as a fitting parameter the results for the column density are similar.} \label{Figure:Fit_NH2_T_5e2Msun_all_wl_T_var_100K_10K_geomspace}
\end{figure*}

Secondly, we consider an exponentially decreasing temperature distribution from 100\,K (at the center) to 20\,K (at the outer edge; see Fig. \ref{Figure:Fit_NH2_T_5e2Msun_all_wl_T_var_100K_10K_geomspace}). 
The qualitative results are similar to those derived considering a linearly decreasing temperature distribution. 
Again, regions with different temperatures can be distinguished from one another.
The column density slope is  derived with a maximum deviation of 30\%, decreasing with distance from the centre (<5\% in the outer regions).
Using the power-law approach for the dust properties, the deviation of the derived column density is $\gtrsim$ 80\% for all cases.\\

\section{Conclusions} \label{Section_Conclusion}
In order to derive information on the column density, temperature and dust emissivity index of molecular clouds and filaments, a usual approach is to apply the modified blackbody fit method in the FIR to sub-millimeter wavelength range.
However, this method is based on multiple assumptions. In this paper we reviewed the resulting basic limitations of this method and evaluate their impact on the derived quantities. The following main conclusions can be drawn:
\begin{itemize}
\itemsep 10pt

\item[1.] By far the highest uncertainty can arise from unknown or poorly constrained optical properties of the dust. In many cases dust opacities are adopted, which were specified by \citet{Hildebrand1983} based on the analysis of observations of a specific reflection nebula. It is not obvious that a general application of this opacity is suitable for the different objects and regions within a given object (characterized by different physical conditions). We therefore propose to first derive the optical depth $\tau$ and subsequently the column density with the aid of a suitable dust model. We find that the optical depth can be derived with high accuracy, especially at longer wavelengths. If the dust properties are known, the BFM provides highly accurate results for the column density $N$(H$_2$) and temperature $T$.

\item[2.] Measurements at four wavelengths are sufficient to obtain accurate results as long as the wavelengths are within $\lambda \in$ (53, 70, 89, 100, 154, 160, 214, 250, 850)\,$\mu$m (SOFIA/HAWC+, Herschel/PACS, Herschel/SPIRE, JCMT/SCUBA-2). In general at least one measurement at a short wavelength (e.g., ALMA band 9, 460\,$\mu$m) is necessary to obtain sufficiently precise values for the column density and temperature for the considered case of a spherical dust distribution with a temperature of 20 K.
.

\item[3.] For very compact, massive objects with optically thick regions the BFM delivers unreliable results. However, molecular clouds and filaments typically resolved with the considered observatories/instruments SOFIA/HAWC+ and Herschel/PACS are optically thin at the considered observing wavelengths on large-scale structures. 

\item[4.] The method yields reliable results for column density and temperature for all dust compositions and grain sizes considered (up to 50$\,\mu$m), with a deviation of less than 10\%, provided the dust properties are known. If a power-law approach for the dust properties (the dust opacity $\kappa_\nu$ is calculated via $\kappa_\nu$ = $\kappa_{\nu_0} (\nu / \nu_0)^\beta$ in this case) is adopted, the column density is under- or overestimated, depending on the chemical composition and grain size (up to 70\%).

\item[5.] The dust emissivity index $\beta_\mathrm{fit}$ is not suitable as an indicator of dust grain size, since on the one hand, the observing wavelengths should be sufficiently long to fulfill the Rayleigh-Jeans approximation. But, on the other hand, at least one short wavelength ($\le$ 460\,$\mu$m) needs to be part of the observing wavelength selection to constrain the location of the maximum of the SED.

\item[6.] If a non-constant temperature distribution is considered, using the BFM it is possible to adequately derive the column density distribution and relative slope of the temperature distribution. Thus, the BFM is particularly well-suited to identify regions with locally embedded heating sources. Compared to clouds with a constant temperature distribution, the column density is traced with an uncertainty of <30\% within the considered parameter space.

\end{itemize}
The BFM allows one deriving reliable constraints on the column density and temperature if the dust properties are well constrained. In contrast to $\beta_\mathrm{SED}$ , the quantity $\beta_\mathrm{fit}$ – derived using the BFM – is not suitable for making statements about the grain size.

\section*{Acknowledgments}
N.Z. and S.W. acknowledge the support by the DLR/BMBF grant 50OR1910.


\subsection*{Conflict of interest statement}

The authors declare no potential conflict of interests.


\section*{ORCID}
\textit{Niko Zielinski} \href{https://orcid.org/0000-0002-3542-2583}{https://orcid.org/0000-0002-3542-2583} \\
\textit{Sebastian Wolf} \href{https://orcid.org/0000-0001-7841-3452}{https://orcid.org/0000-0001-7841-3452} \\


\clearpage
\clearpage
\bibliography{lit}%

\end{document}